# Generating 10-GHz phonons in nanostructured silicon membrane optomechanical cavity

J. Zhang[1,*], X. Le Roux[1], M. Montesinos-Ballester[1], O. Ortiz[1], D. Marris-Morini[1], E. Cassan[1], L. Vivien[1], N. D. Lanzillotti-Kimura[1], and C. Alonso-Ramos[1]
[1] Université Paris-Saclay, CNRS, Centre de Nanosciences et de Nanotechnologies, 91120 Palaiseau, France
jianhao.zhang@c2n.upsaclay.fr

Flexible control of photons and phonons in silicon nanophotonic waveguides is a key feature for emerging applications in communications, sensing and quantum technologies. Strong phonon leakage towards the silica under-cladding hampers optomechanical interactions in silicon-on-insulator. This limitation has been circumvented by totally or partially removing the silica under-cladding to form pedestal or silicon membrane waveguides. Remarkable optomechanical interactions have been demonstrated in silicon using pedestal strips, membrane ribs, and photonic/phononic crystal membrane waveguides. Still, the mechanical frequencies are limited to the 1-5 GHz range. Here, we exploit the periodic nanostructuration in Si membrane gratings to shape GHz phononic modes and near-infrared photonic modes, achieving ultrahigh mechanical frequency (10 GHz) and strong photon-phonon overlap (61.5%) simultaneously. Based on this concept, we experimentally demonstrate a one-dimension optomechanical micro-resonator with a high mechanical frequency of 10 GHz and a quality factor of 1000. These results were obtained at room temperature and ambient conditions with an intracavity optical power below 1 mW, illustrating the efficient optical driving of the mechanical mode enabled by the proposed approach.

**Introduction**

On-chip optomechanical interactions have been intensively studied in the context of cavity optomechanics [1-3] and Brillouin scattering [4], using a wide variety of photonic technologies, including chalcogenides materials [5], gallium arsenide [6-9], aluminum nitride [10], and lithium niobate [11], to name a few. Optomechanical interactions in silicon have long been precluded by strong phonon leakage towards the silica cladding [12-14]. This limitation has been recently circumvented by mechanically isolating the silicon waveguide core from the substrate through partial [15] or total removal of the silica under-cladding [16]. This approach revolutionized the field of silicon optomechanics, allowing the demonstration of a myriad of optomechanical features and functionalities [17-19], targeting applications in communications, sensing, and quantum technologies. Different geometries have been proposed and demonstrated to control photons and phonons in silicon photonic waveguides. Most remarkable results have been achieved using pedestal strips [15], membrane ribs [16-18], and membrane photonic/phononic crystal waveguides [20-28]. However, the mechanical frequencies of these optomechanical cavities were limited to the 1-5 GHz range. Achieving higher frequencies, e.g., in the microwave X band (8-12 GHz), could be very interesting for radiofrequency signal generation and processing in radar and communication applications [14].

Here, we propose and experimentally demonstrate optomechanical silicon membrane gratings with high-frequency mechanical modes (> 5GHz) and a strong photon-phonon overlap. We harness Si periodic nanostructuration to increase the frequency of the mechanical mode while shaping the field distribution of the optical mode to maximize optomechanical coupling. We demonstrate a silicon optomechanical resonator with high mechanical frequency and quality factors of 10 GHz and 1000, respectively. The phononic mode is driven by a sideband-unresolved optical resonance, with intracavity power below 1 mW.

## Results
**Optomechanical silicon membrane grating.**

Figure 1(a) shows a schematic view of the proposed nanostructured silicon optomechanical membrane grating, comprising a waveguide core of width W and a periodic corrugation defined by the pitch Λ, and the length ($L_T$) and width ($W_T$) of the teeth. We set silicon thickness to $t_{Si}$ = 220 nm for compatibility with standard Si photonics. In Fig. 1(b), we show the calculated mechanical frequency as a function of the waveguide width (W) for Λ = 420 nm, $L_T$ = 80 nm, and $W_T$ = 400 nm. The mechanical frequency can be tuned between 9 GHz and 18 GHz with the deformation of mechanical mode concentrated in the waveguide core, just by changing the waveguide width. We choose a width of W = 340 nm as a compromise between high mechanical frequency and good mechanical and optical confinement. Figure 1(c) plots the mechanical band diagram of the proposed waveguide. Due to phase-matching and energy conservation conditions, effective optomechanical backaction [2,3] requires short mechanical wavevectors ($q_z$~0). In Fig. 1(d), we plot the mechanical mode's calculated displacement profile near $q_z$=0. Most of the displacement is confined in the narrow silicon section between the grating teeth. Indeed, the mode's high frequency is directly related to its tight confinement within a small silicon volume ( ~ 220 nm × 340 nm × 250 nm).

The optical band diagram of the proposed waveguide is shown in Fig. 1(e). The waveguide supports two guided optical modes below the light line. One lies below the photonic bandgap, and the other above. These modes are referred to here as L-band mode and U-band mode, respectively. The modal profiles of the L and U band modes are shown in Figs 1(f) and 1(g). The L-band mode is mostly confined within the grating teeth, while the U-band mode is confined in the gaps between the teeth. The different field confinements strongly affect the optomechanical backaction, which is related to the overlap between the mechanical and optical modes [29]. The L-band mode yields an overlap with the mechanical mode of only 16.8%, while the U band mode achieves a strong overlap of 61.7%. The large overlap allows a strong optomechanical coupling rate [30-35] as large as $g_o$=210 kHz. Hence, the proposed Si grating membrane stands as a promising candidate for high-frequency optomechanics with strong optomechanical coupling.

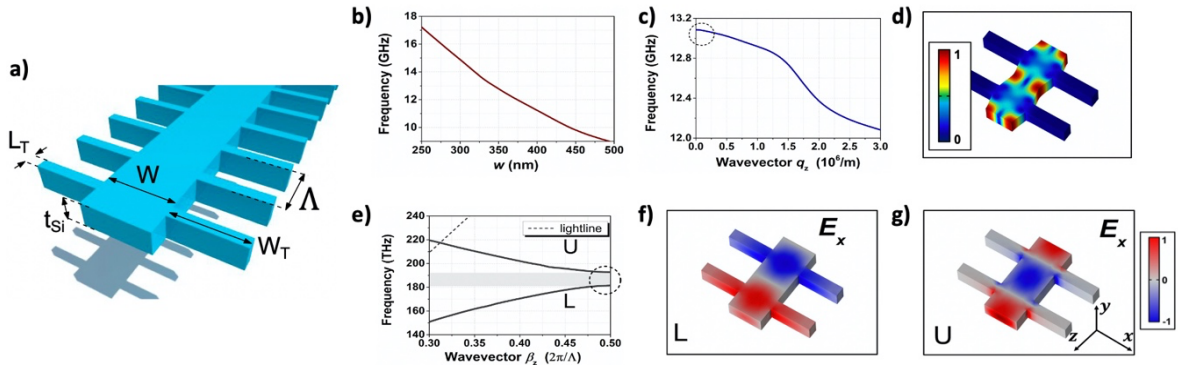

**Figure 1: a)** Schematic view of proposed optomechanical Si membrane grating. **b)** Mechanical frequency calculated as a function of the waveguide width for Λ = 420 nm, $L_T$ = 80 nm, and $W_T$ = 400 nm. **c)** Mechanical band diagram and **d)** normalized displacement profile of proposed optomechanical mode. **e)** Optical band diagram of proposed Si membrane grating. Optical profile distribution of **f)** U-band mode and **g)** L-band mode.

**Probing the 10-GHz phonons in optomechanical crystal cavity**

To demonstrate the proposed concept, we developed an optomechanical micro-resonator. The cavity is formed by parabolically reducing the lattice period, Λ, [36]. The period in the center is 420 nm. The period in the cavity sides is 380nm. The waveguide core width is W = 340 nm. The teeth length and width are $L_T$ = 80 nm and $W_T$ = 400 nm. We use silicon membrane waveguides, with subwavelength periodic cladding [37], to connect the input and output of the cavity with fiber-chip grating couplers. This approach allows copropagating optical excitation and readout of the cavity response using cleaved SMF-28 fibers and integrated access waveguides. We fabricated the optomechanical Si grating cavity using electron-beam lithography and reactive ion etching. The silica under-cladding was removed by selective etching with vapor hydrofluoric acid. Figure 2(a) shows the scanning electron microscope (SEM) images of the fabricated microresonator.

Figure 2(b) depicts the experimental characterization setup. The optical reflection and transmission are monitored with an optical/electronic spectrum analyzer (OSA/ESA). Experimental measurements were performed at room temperature and at atmospheric pressure. The mechanical motion and optomechanical backaction are excited using the first-order optical mode (U band mode), with a measured resonance wavelength of 1538 nm and optical quality factor of 7730 (FWHM 0.199 nm). This quality factor allows sideband unresolved optomechanical interaction. The optomechanical crystal is driven using a tunable laser, with 0.316 mW (-5 dBm) of optical power coupled into the access silicon waveguide. The pump wavelength is scanned to approach the resonance from a blue-detuned position which stiffens the optomechanical resonator [2,3]. Due to thermo-optic and nonlinear effects in silicon, the cavity resonance drifts dynamically with the increasing intra-cavity power. As shown in Fig. 2(c), the optomechanical cavity exhibits one oscillating frequency near 10 GHz. This mechanical mode yields a mechanical quality factor of 1110 (see Fig. 2(d)). The optical pump is blue detuned from the optical cold resonance wavelength by 1.14 nm, with an estimated intracavity power of only 814 µW. The measured frequency (10.14 GHz) is slightly lower than the calculated for a grating with a uniform period (Fig. 1(b) and 1(c)). This is expected, as the period chirping required to form the cavity affects the mechanical mode distribution and thus the overall frequency of the cavity. Fabrication imperfections can also shift the measured frequency compared to simulations. This 10 GHz mechanical frequency is a substantial increase compared with state-of-the-art silicon one-dimension optomechanical photonic/phononic crystal cavities, with a typical frequency near 5 GHz [21, 24].

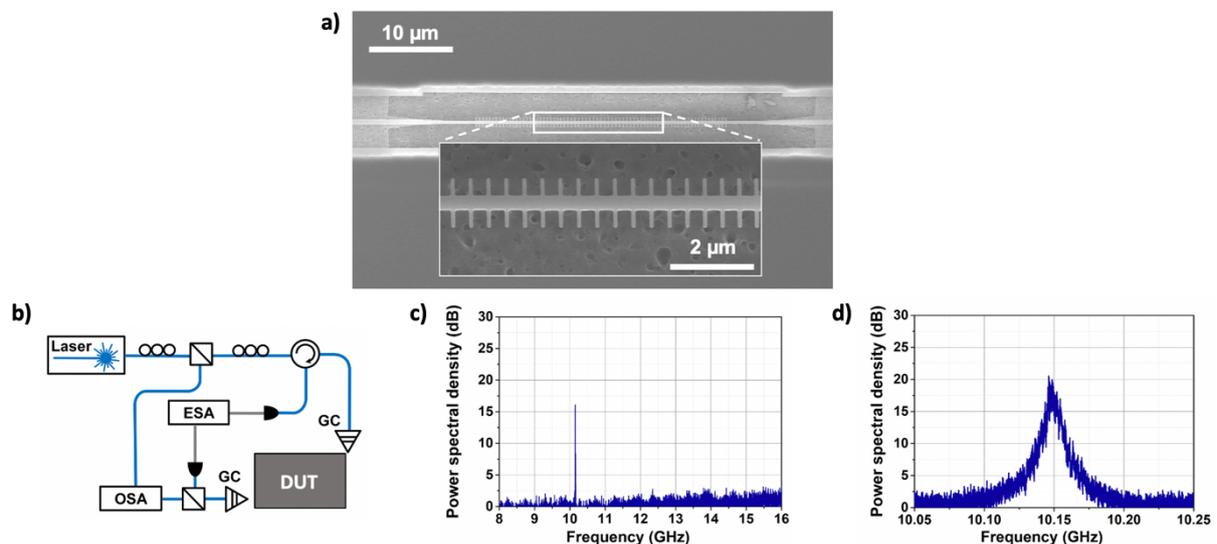

**Figure 2: a)** Scanning electron microscope images of the fabricated optomechanical micro-resonator. b) The setup used for optomechanical characterization. ESA: electronic spectrum analyzer. OSA: optical spectrum analyzer. GC: grating coupler. DUT: device under test. c) and d) radio-frequency spectrum of the optomechanical crystal, driven by an optical power of 316 µW in the access waveguide.

## Conclusions

In summary, we have proposed and demonstrated a new strategy to confine high-frequency phonons and near-IR photons in silicon membrane microresonators. We leverage phononic mode profile engineering in periodically nanostructured silicon membrane gratings to confine the mechanical mode in narrow silicon core between grating teeth, thereby achieving high mechanical frequency (10 GHz). Simultaneously, periodic nanostructuration provides optical confinement control that we exploit to achieve strong photon-phonon overlap (61.7%). We demonstrate this concept by implementing an optomechanical micro-resonator with a measured mechanical frequency of 10 GHz. This is a high-frequency record for one-dimension silicon phononic/phononic crystal cavities [2, 21]. These results open new perspectives to exploit high-frequency phonons in silicon photonics, based on periodic silicon nanostructuration, with great potential for communications and sensing applications.